\documentclass{ws-ijmpe}
\usepackage[super,compress]{cite}
\usepackage{psfig}
\begin{document}

\markboth{N. Paar, T. Marketin, D. Vale, and D. Vretenar}{Modeling nuclear weak-interaction processes with relativistic energy density functionals}

\catchline{}{}{}{}{}

\title{MODELING NUCLEAR WEAK-INTERACTION PROCESSES WITH RELATIVISTIC 
ENERGY DENSITY FUNCTIONALS}
\author{N. PAAR}
\address{Department of Physics, University of Basel, Klingelbergstrasse 82, CH-4056 Basel, Switzerland\\
Department of Physics, Faculty of Science, University of Zagreb, Croatia}
\author{ T. MARKETIN, D. VALE, D. VRETENAR}
\address{Department of Physics, Faculty of Science, University of Zagreb, 
Croatia}

\maketitle

\begin{history}
\received{Day Month Year}
\revised{Day Month Year}
\end{history}

\begin{abstract}
Relativistic energy density functionals have become a standard framework for nuclear structure studies of ground-state properties and collective excitations over the entire nuclide chart. We review recent developments in modeling nuclear weak-interaction processes: charge-exchange excitations and the role of isoscalar proton-neutron pairing, charged-current neutrino-nucleus reactions relevant for supernova evolution and neutrino detectors, and calculation of $\beta$-decay rates for r-process nucleosynthesis.
\end{abstract}

\keywords{nuclear weak interactions; neutrino-nucleus reactions; beta-decay rates}

\ccode{PACS numbers: 21.30.Fe,21.60.Jz,23.40.Bw,25.30.Pt}


\section{\label{sec1}Introduction}

Nuclear weak-interaction processes play a crucial role during various phases of supernova evolution and in the associated production of chemical elements~\cite{Lan.03,Lan.11}. In the presupernova phase, electron capture and $\beta$-decays produce neutrinos and change the number of electrons in a way that directly influences the collapse dynamics and subsequent explosion~\cite{Jan.07}. The r-process nucleosynthesis, 
responsible for the synthesis of about half the elements heavier than iron, is governed by 
neutron capture and $\beta$-decays. Neutrino-induced reactions on nuclei also play an important
role in supernova environment. Neutral-current neutrino-nucleus scattering in stellar environment determines the rate of cooling 
by neutrino transport~\cite{Woo.90,Hax.88}. Inelastic neutrino-nucleus scattering has been included in supernova simulations 
as a novel mode of energy exchange between neutrinos and matter
~\cite{Bru.91,Lan.08,Lan.10}. Furthermore, charged-current 
neutrino-nucleus reactions are important in the environment of exploding massive stars,
considered as one of the possible sites for r-process nucleosynthesis. Neutrino-induced reactions in explosive
supernova nucleosynthesis significantly contribute to the production of not only radioactive, but also
long-lived nuclides~\cite{Sie.15}.

As emphasized in Ref.~\refcite{Jan.07}, a real breakthrough in our understanding of the inner workings of supernova explosions, 
based on self-consistent models with all the relevant microphysics included, has not been achieved yet. 
The necessary nuclear input is obtained either from a variety of independent sources based on different effective
interactions and approximations, or it is based on relatively crude estimates that cannot provide a realistic 
description of key weak-interaction processes~\cite{Bru.85}.
Self-consistent modeling of heavy-element nucleosynthesis presents a particularly difficult challenge, 
both because of the nature of the r-process and the large amount of nuclear data input involved. 
As the detailed description of this process requires a knowledge of the structure and decay properties of thousands of nuclei, 
most of which are outside the reach of present-day experimental facilities, it is necessary 
to develop a universal self-consistent theoretical framework that can systematically be used over the 
entire nuclear chart. One of the key phenomena involved 
is $\beta$-decay, and the closely related $\beta$-delayed neutron emission, as it determines the
rate of the r-process and directly influences the final elemental abundances. We note, however, 
that there are few systematic studies of the impact of first-forbidden transitions on the total decay
rates of neutron-rich nuclei.

In this article we present an overview of current developments in self-consistent modeling of
weak-interaction processes and associated nuclear transitions based on the framework of relativistic 
nuclear energy density functionals (RNEDF).
The focus is on charged-current neutrino-nucleus reactions and $\beta$-decays. Many results of alternative theoretical approaches to $\beta$-decay 
rates, neutrino-nucleus reactions and their role in supernova dynamics and nucleosynthesis have been reviewed in Refs.~\refcite{Lan.03,Lan.11,Jan.07,Bor.06,Bal.15} and references therein. Over the last decade significant progress has been achieved in modeling ground-state nuclear properties, excitations and associated weak-interaction phenomena using the RNEDF framework. Recent studies include the analysis of collective nuclear phenomena, including giant resonances and exotic modes of excitation~\cite{PVKC.07,PNVM.09,Kha.11,Vre.12}, and their application in constraining  the neutron skin of neutron-rich nuclei~\cite{Vre.03,Kli.07,Pie.12,Kra.13}, symmetry energy and neutron star properties~\cite{Paa.14}. RNEDF-based models have also been used in calculations of transition matrix elements that contribute to a variety of weak-interaction processes, including $\beta$-decay of r-process nuclei~\cite{Nik.05,Mar.07}, muon capture~\cite{Mar.09} and stellar electron capture~\cite{Niu.11}, neutral-current~\cite{Dap.11} and charged-current neutrino-nucleus reactions~\cite{Paa.08,Paa.11,Sam.11,Paa.13}. The investigation of charge-exchange transitions in nuclei and  comparison with available data provide important benchmark tests for microscopic models of weak-interaction processes~\cite{Paa.11,Mar.12,MLVR.12,Lit.14}.  

The paper is organized as follows. The theory framework is briefly outlined in Sec.~\ref{sec2}. In Sec.~\ref{sec3} we present a benchmark test for charge-exchange excitations in $^{56}$Fe. Sec.~\ref{sec4} explores the effect of isoscalar pairing interactions on Gamow-Teller transitions in light and  medium-mass nuclei. In Sec.~\ref{sec5} charged-current neutrino-nucleus cross sections are considered, and in Sec.~\ref{sec6} results of large-scale calculations are analyzed. $\beta$-decay half-lives and results of systematic calculations 
are discussed in Section~\ref{sec7}. Sec.~\ref{sec8} includes a brief summary and concluding remarks. 

\section{\label{sec2}Theory framework}

The analysis of collective excitations and nuclear weak-interaction processes is based on the framework of relativistic nuclear energy density functionals (RNEDFs). At the mean-field level ground-state calculations of open-shell nuclei are implemented using the relativistic Hartree-Bogoliubov (RHB) model~\cite{Vre.05}, based on density-dependent meson-nucleon effective interactions~\cite{Nik.02}. A detailed descripton of the formalism and the corresponding computer codes are included in Ref.~\cite{Nik.14}. In this work we employ the relativistic functional DD-ME2~\cite{Lal.05} in the particle-hole channel, and pairing correlations in open-shell nuclei are described by the finite-range Gogny interaction D1S~\cite{Ber.91}. In the small-amplitude limit collective nuclear excitations relevant for weak-interaction processes are calculated using the relativistic quasiparticle random-phase approximation (RQRPA)~\cite{PVKC.07}. The RQRPA is formulated in the canonical single-nucleon basis of the RHB model, and the residual interaction is derived from the same relativistic functional used to compute ground state properties. An important advantage of this framework is that it employs in a consistent way the same effective interaction both in the RHB equations that determine the canonical basis, and in the matrix equations of the RQRPA. 

\section{\label{sec3}Charge-exchange excitations}

\begin{figure}[th]
\centerline{\psfig{file=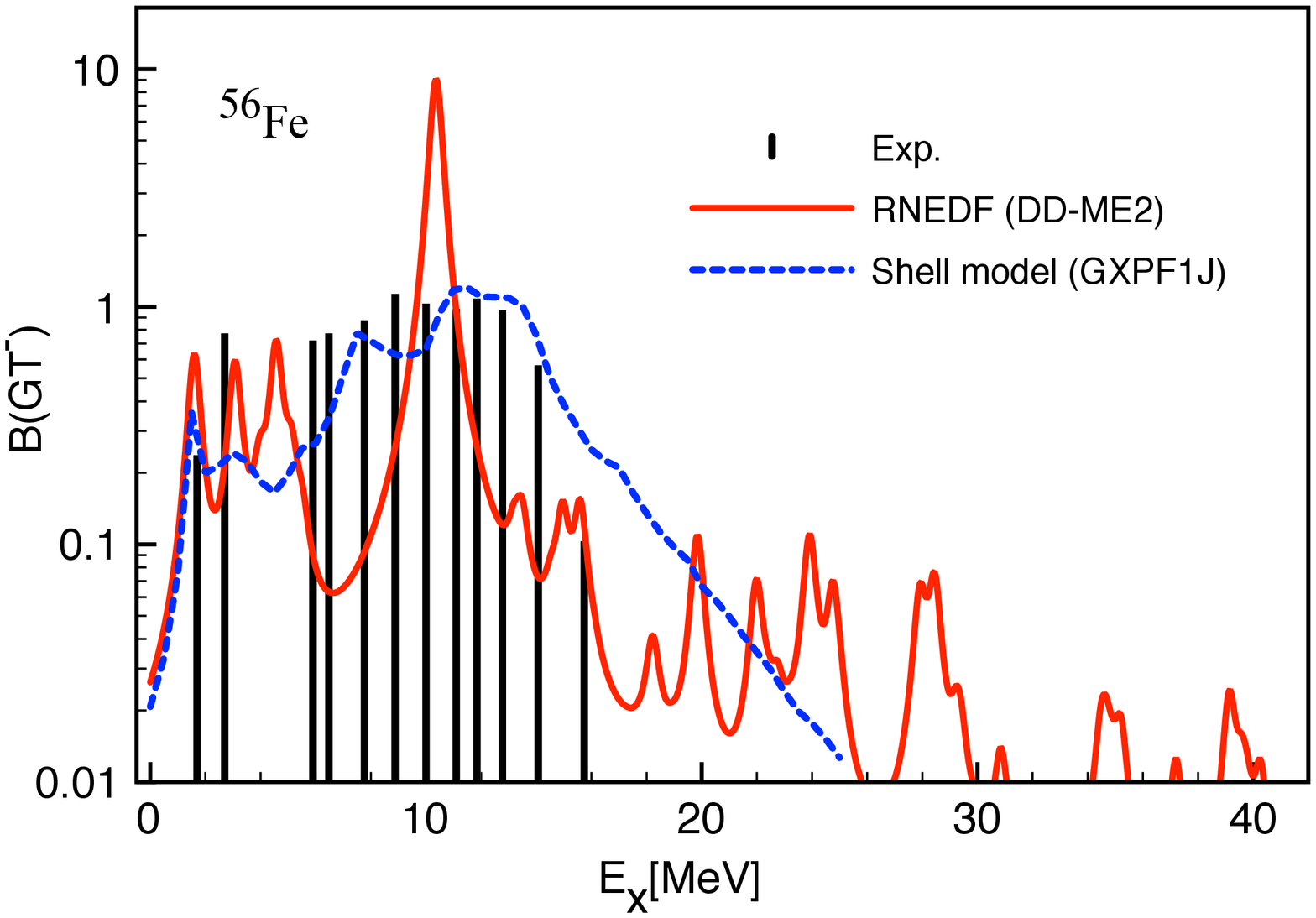,width=9cm,angle=-90}}
\vspace{-1.5cm}
\caption{The Gamow-Teller (GT$^-$) transition strength distribution for $^{56}$Fe,
shown as a function of excitation energy in the final nucleus. The RQRPA results 
based on the RNEDF DD-ME2 are compared to the
shell model calculations (GXPF1J)~\cite{Paa.11}, and available data from $(p,n)$
reactions~\cite{Rap.83}.}
\label{Fe56_gt_qrpa_shell_exp}
\end{figure}

Gamow-Teller(GT$^\pm$) excitations in nuclei correspond to spin-flip and isospin-flip transitions, involving both the spin and isospin operators $\sigma\tau_{\pm}$. Recently GT$^{\pm}$ transitions in $^{54,56}$Fe and $^{58,60}$Ni have been explored in a comparative study based on the RNEDFs, Skyrme energy density functionals, and the shell model~\cite{Paa.11}. As a benchmark case, Fig.~\ref{Fe56_gt_qrpa_shell_exp} displays the GT$^-$ strength distributions for $^{56}$Fe, computed using two different theory frameworks: the relativistic QRPA based on the functional DD-ME2 and the shell model (GXPF1J). The data from $(p,n)$ reactions~\cite{Rap.83} are also included for comparison. The calculated transition strengths are folded by a Lorentzian function of width $\Gamma$=0.5 MeV. More details about the shell model Hamiltonian GXPF1J and its implementation in the $pf$ shell nuclei are given in Refs.~\cite{Hon.02,TS.09}. Although the QRPA model calculation includes only $2qp$ configurations and, therefore, cannot reproduce the detailed structure of excitation spectra obtained by the shell model, one finds a reasonable agreement with the global features of the GT$^{-}$ transition strength distribution. We note, however, that the measured overall GT$^{\pm}$ transition strengths in  $^{54,56}$Fe and $^{58,60}$Ni can only be reproduced  by quenching the free-nucleon axial-vector coupling constant $g_A$~\cite{Paa.11}. This is true for all models and the respective parameterizations employed in the study: i) RNEDF (DD-ME2), RPA based on Skyrme functionals (SGII, SLy5) and iii) the shell model (GXPF1J)~\cite{TS.09}. The quasiparticle RPA calculations include the quenching of the free-nucleon axial-vector coupling constant $g_A = 1.262 \to g_A = 1$, corresponding to the quenching factor 0.8 in the GT transition operator. A quenching factor 0.74, used in shell model calculations, is adapted to the effective interaction and model space under consideration~\cite{TS.09}. In the case of $^ {56}$Fe, the RNEDF QRPA calculation with $g_A = 1$ yields the B(GT$^-$) within experimental uncertainties~\cite{Paa.11}. The quenching of $g_A$, however, should be considered with caution because of the well known problem of missing GT strength, either due to excitations that involve complex configurations at higher excitation energies~\cite{Ber.82}, or excitations that include non-nucleonic degrees of freedom~\cite{Ari.01}.

\section{\label{sec4}Gamow-Teller transitions and the isoscalar pairing interaction}

Microscopic descriptions of open-shell nuclei based on energy density functionals
necessitate the inclusion of an effective pairing interaction. The isovector $(T=1,S=0)$ part
of of this interaction accounts for the odd-even staggering in the separation 
energies, and for the gap in the excitation spectrum of even-even 
nuclei~\cite{Boh.58,Boh.69,Bri.05}. In the relativistic framework, standard 
mean-field methods that include pairing correlations in the description of
ground-state properties of open-shell nuclei are the relativistic Hartree + 
Bardeen-Cooper-Schrieffer (BCS) model~\cite{Rin.96}, and relativistic Hartree-Bogoliubov (RHB) 
model~\cite{Vre.05}. A fully self-consistent calculation of excitations in open-shell 
nuclei involves the inclusion of particle-particle correlations not only at the 
ground-state level, but also in the residual interaction of the 
quasiparticle RPA equations~\cite{Paa.03}. In the case of
charge-exchange transitions, the particle-particle interaction 
$V^{pp}$ includes not only the usual isovector ($T=1$) part, which has the same
form as the pairing interaction used for ground-state calculation (e.g., the Gogny pairing), 
but also the isoscalar $T=0$ channel, for which one could adopt the
following form:

\begin{equation}
V_{12}
= - V_0 \sum_{j=1}^2 g_j \; {\rm e}^{-r_{12}^2/\mu_j^2} \;
    \hat\Pi_{S=1,T=0} \, ,
\label{pn-pair}
\end{equation}
where $\hat\Pi_{S=1,T=0}$ projects onto states with $S=1$ and $T=0$. The standard 
choice for the ranges of the two
Gaussians is $\mu_1$=1.2\,fm and $\mu_2$=0.7\,fm, together with the 
relative strengths $g_1 =1$ and
$g_2 = -2$ so that the force is repulsive at small distances~\cite{Paa.04}.
The remaining free parameter is $V_0$, the overall strength that can only be
constrained by data on nuclear excitations and/or decays.
As pointed out in several HFB+QRPA calculations, by adjusting the strength parameter
of isoscalar pairing interaction to selected data, one can improve the description of 
$\beta$-decay half-lives in neutron-rich nuclei relevant for the 
r-process nucleosynthesis (see Sec.~\ref{sec7} for more details)~\cite{Eng.99,Bor.00,Yos.13}.

In the study of Ref.~\refcite{Paa.04} it has been shown that the $T=0$ particle-particle interaction
leads to considerable modification of the Gamow-Teller strength distribution
in Sn isotopes. In particular, the energy spacings between the main GT resonance peaks
depend on the strength of $T=0$ pairing, and the low-energy GT transition strength 
becomes more pronounced by increasing $V_0$. Because of the complex dependence of 
GT transition strength on the $T=0$ pairing strength in various mass regions, 
it is difficult to determine a unique global value of $V_0$ from available experimental results. 
Since the modeling of $\beta$-decay rates  
crucially depends on the low-energy part of GT excitation spectra~\cite{Nik.05}, a possible 
approach is to employ experimental $\beta$-decay half-lives to constrain $V_0$ locally for
specific groups of nuclei, or to introduce a smooth mass dependence for the strength
parameter (Sec.~\ref{sec7}). 

The role of $T=0$ pairing for GT states in $N=Z$ ($A=48-64$) nuclei was investigated using 
the Hartree-Fock-Bogoliubov + QRPA model based on Skyrme functionals~\cite{Bai.13}. 
It has been shown that the low-energy excitation strength is enhanced
by $T=0$ pairing and that $N=Z$ nuclei could provide important information about 
this channel of the effective interaction in medium-heavy and heavy mass nuclei~\cite{Bai.13}. 
Recently the GT transition strength distributions were explored in $(^{3}He,t)$ 
charge-exchange reactions with f-shell nuclei~\cite{Fuj.14}, resulting in a dominant concentration 
of GT strength in low-energy states, in particular for $^{42}$Ca, $^{46}$Ti, and $^{50}$Cr.
As shown in Refs.~\refcite{Bai.13,Bai.14}, the isoscalar spin-triplet particle-particle interaction
mixes largely the $(\nu j_{>}=l+1/2 \to \pi j_{<}=l-1/2)$ configurations into the low-energy states, and this 
mixing plays an important role in the formation and evolution of collectivity of low-energy GT states.
\begin{figure}[th]
\centerline{\psfig{file=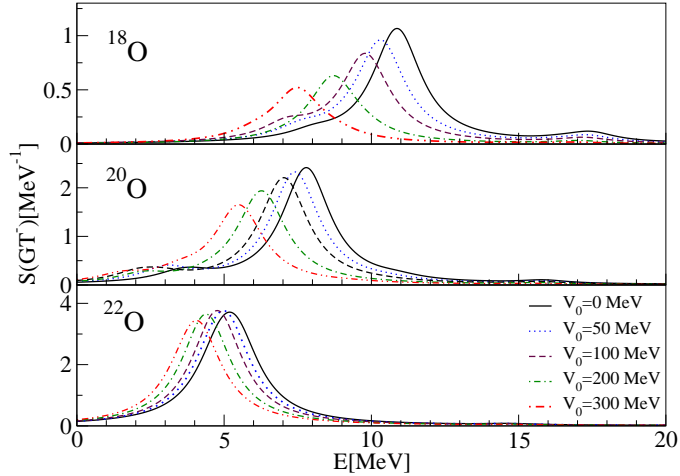,width=7.5cm,angle=-90}}
\caption{Gamow-Teller (GT$^-$) strength distributions 
for $^{18,20,22}$O calculated with the RHB+RQRPA model (DD-ME2 functional and 
Gogny pairing in the 
$T=1$ channel). The strength of the 
$T=0$ particle-particle
interaction Eq.~(\ref{pn-pair}) varies from $V_0 =0$ to $V_0=300$ MeV.}
\label{gtm_O18_O22}
\end{figure}

In Fig. ~\ref{gtm_O18_O22} we plot the RHB+RQRPA 
GT$^-$ transition strength distributions for $^{18,20,22}$O. The self-consistent 
ground states are computed using the RHB model (DD-ME2 functional and 
Gogny pairing in the $T=1$ channel), and the transition
strengths are obtained in the relativistic QRPA that includes 
both the $T=1$ and $T=0$ pairing interactions. In the $T=0$ case,
the strength parameter is varied in 
the range of values from $V_0 =0$  to $V_0 =300$ MeV.
For $^{18}$O the main GT$^-$ state at 10.8 MeV ($V_0$=0 MeV) corresponds
predominantly to the $\nu 1d_{5/2} \to \pi 1d_{3/2}$ transition. However, by increasing the
strength of the $T=0$ pairing interaction the structure of the main GT$^-$ peak 
becomes more complex. For example, for  $V_0$=300 MeV the main peak is shifted 
to 7.5 MeV, and the dominant configurations $\nu 1d_{5/2} \to \pi 1d_{3/2}$,
$\nu 1d_{3/2} \to \pi 1d_{5/2}$, $\nu 2s_{1/2} \to \pi 1d_{3/2}$, 
$\nu 1d_{3/2} \to \pi 2s_{1/2}$, $\nu 2s_{1/2} \to \pi 2s_{1/2}$, and  $\nu 1d_{5/2} \to \pi 1d_{5/2}$, 
contribute with 65$\%$, 10$\%$, 10$\%$, 6$\%$, 5$\%$, and 1$\%$, respectively, to the 
QRPA amplitude $X^2-Y^2$. We note that for all the additional configuration pairs except the last one, 
the occupation probabilities of both single-particle states are rather small, 
i.e. the contribution of these particle-like 
configurations is due to the $T=0$ pairing interaction in the QRPA. Because of destructive
interference between the $\nu 1d_{5/2} \to \pi 1d_{3/2}$ and particle-like 
configurations, the overall GT$^-$ strength is reduced for larger values of $V_0$.
As shown in Fig.~\ref{gtm_O18_O22} for the oxygen isotopes, the GT$^-$ transition
strength is sensitive to the neutron excess: from $^{18}$O to $^{22}$O
the main peak is shifted to lower energy by 5.4 MeV. While for $^{18}$O ($V_0=0$ MeV)
the structure of the main peak is dominated by a single transition, for $^{22}$O
the QRPA amplitude of the main peak at 5.42 MeV is almost equally distributed 
between two configurations: $\nu 1d_{5/2} \to \pi 1d_{3/2}$ and 
$\nu 1d_{3/2} \to \pi 1d_{3/2}$. 

\begin{figure}[th]
\centerline{\psfig{file=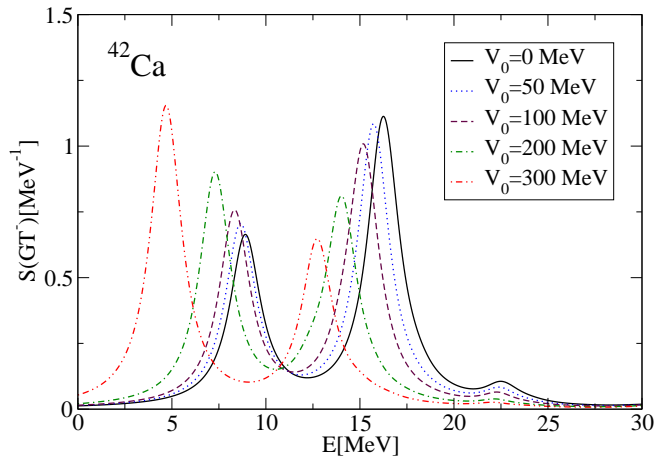,width=7.5cm,angle=-90}}
\caption{Same as the caption to Fig.~\ref{gtm_O18_O22}, but for $^{42}$Ca.}
\label{gtm_Ca42}
\end{figure}

Figure~\ref{gtm_Ca42} displays the calculated GT$^-$ strength distributions for
$^{42}$Ca. Similar to the previous case, the dependence of transition spectra 
on the $T=0$ particle-particle interaction Eq.~(\ref{pn-pair}) is explored 
by varying the pairing strength. In the
case without isoscalar pairing ($V_0=0$), one finds a dominant peak 
at 16.2 MeV that corresponds to the transition $\nu 1f_{7/2} \to \pi 1f_{5/2}$. Another
prominent state is calculated in the low-energy region at 8.9 MeV 
($\nu 1f_{7/2} \to \pi 1f_{7/2}$). For $V_0=0$ MeV, the
B(GT) values for the main transitions $(\nu j_{>} \to \pi j_{>})$ and  $(\nu j_{>} \to \pi j_{<})$ exhaust  
34$\%$ and 53$\%$ of the GT sum rule $3(N-Z)=6$, respectively.
However, recent data from $(^{3}He,t)$ reactions indicate that 80$\%$
of the GT strength is carried by the lowest GT state~\cite{Fuj.14}. The 
$T=0$ particle-particle interaction provides a possible
mechanism to shift the calculated strength toward lower excitation 
energies~\cite{Fuj.14,Bai.14}. This effect is demonstrated in Fig.~\ref{gtm_Ca42} 
for the range of values of the $T=0$ pairing strength parameter $V_0=50-300$ MeV.
By increasing $V_0$ the peaks are lowered in energy, the B(GT) value of
the low-energy state is considerably enhanced, whereas the peak of the high-energy
state decreases. For $V_0=300$ MeV the low-lying state exhausts 59$\%$ 
and the high-lying state 33$\%$ of the GT sum rule.
The origin of the discrepancy with experimental results, which show no evidence of  
pronounced high-energy GT transitions~\cite{Fuj.14}, remains
an open problem. As pointed out in Ref.~\refcite{Bai.14}, an additional effect 
could be due to contributions of the tensor force. However, in the case
of $^{42}$Ca this effect is negligible in the low-energy states, whereas the high-energy
states are shifted downward in energy by 1-2 MeV~\cite{Bai.14}.

\section{\label{sec5}Charged-current neutrino-nucleus reactions}

In charged-current neutrino-nucleus reactions an incoming electron neutrino $(\nu_e)$
induces a charge-exchange in the target nucleus $X(Z,N)$,
\begin{equation}
\nu_e + \, _{Z}X _N  \rightarrow \, _{Z+1}X^*_{N-1} + e^- \;.
\end{equation}
The general formalism for calculating neutrino-nucleus cross sections, derived using the 
weak-interaction Hamiltonian in the current-current form, is given in detail in 
Refs.~\refcite{Con.72,Wal.75}. The cross section includes the transition matrix elements  
of the charge $\hat{\mathcal{M}}_J$, longitudinal $\hat{\mathcal{L}}_J$, 
transverse electric $ \hat{\mathcal{T}}_J^{EL}$, and transverse magnetic 
$\hat{\mathcal{T}}_J^{MAG}$ multipole operators, between the initial $|J_i \rangle$ and 
final $|J_f \rangle$ nuclear states \cite{Con.72}. A
complete description of inelastic neutrino-nucleus scattering in the energy 
range of supernova neutrinos requires the inclusion of higher multipoles $J$~\cite{Paa.08}. 
Although the contribution from higher-order multipole transitions to the cross section is rather small at 
low incoming neutrino energies, these cannot be neglected at neutrino energies of tens of MeV. 

\begin{figure}[]
\centerline{\psfig{file=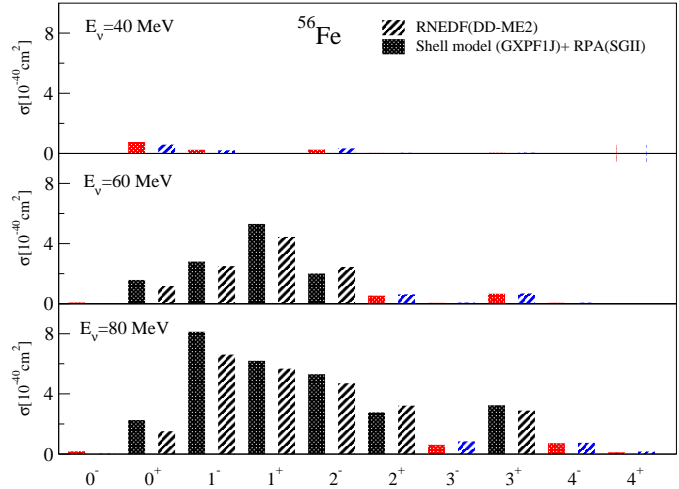,width=7.5cm,angle=-90}}
\caption{Contributions of the multipole transitions $J^{\pi}=0^{\pm} - 4^{\pm}$ to the 
inclusive cross section for the $^{56}$Fe($\nu_e,e^-$)$^{56}$Co reaction,
at $E_{\nu_e}$ = 40, 60 and 80 MeV.
The results correspond to RHB+RQRPA (DD-ME2) calculations, and the shell model (GXPF1J) (for the $1^+$ 
transition) plus the RPA (SGII) for higher multipoles~\cite{Paa.11}.}
\label{FE56_shell_vs_rpa}
\end{figure}

When interacting with nuclei, neutrinos induce transitions from the initial
state of a target nucleus (i.e. the ground state $J^{\pi}$=0$^+$) to a large 
number of excited states that are allowed by conservation of energy and
selection rules. The contribution of various multipole transitions to the 
neutrino-nucleus interaction cross sections essentially depends 
on the energy of incoming neutrinos. Figure ~\ref{FE56_shell_vs_rpa} 
displays the contributions of the multipole
transitions $J^{\pi}=0^{\pm} - 4^{\pm}$ to the inclusive cross section for
the $^{56}$Fe($\nu_e,e^-$)$^{56}$Co reaction, at selected neutrino energies
$E_{\nu_e}$ = 40, 60 and 80 MeV. In the  
RHB+RQRPA (DD-ME2) model calculation  
the value of axial-vector coupling constant $g_A$=1 is used for all multipole 
operators. The partial contributions are compared with those of 
hybrid calculation that employs the shell model 
(GXPF1J) for the $1^+$ channel, whereas the RPA based on the Skyrme
functional SGII is used for higher multipoles~\cite{Paa.11}. 
For lower neutrino energies, $E_{\nu} \lesssim$ 40 MeV, the dominant contribution to the 
cross section originates from $J^{\pi}$=1$^+$ transitions. However, with increasing 
neutrino energy higher multipole transitions start contributing
to the cross section, as shown for $E_{\nu}$=60 MeV and 80 MeV. We note that 
the two approaches predict a similar structure and distribution of the relative contributions 
from various multipoles.

Experimental results for inelastic neutrino-nucleus cross sections are rather limited. 
For two target nuclei $^{12}$C and $^{56}$Fe, data are available from the 
LAMPF~\cite{Kra.92}, KARMEN~\cite{Bod.94} and LSND~\cite{Ath.97,LSND.98,LSND.01} 
collaborations. To compare model calculations with the data, 
cross sections are averaged over fluxes which depend on the neutrino
source. In the case of  $\nu_e$, the Michel flux from the decay at rest~(DAR) of muons ($\mu^+$)
is used \cite{Kra.92}: 
$
f(E_{\nu_e}) = 
 ({96 E_{\nu_e}^2}/{m_{\mu}^4})  (  m_{\mu} -2
E_{\nu_e}  ).
$
The flux averaged cross section for the $^{12}$C target has been calculated 
using nonrelativistic~\cite{Aue.97,VAC.00} and 
relativistic~\cite{Paa.08,Sam.11,Paa.13} energy density functionals, 
the projected quasiparticle
random phase approximation~\cite{Sam.11}, and the shell 
model~\cite{Hay.00,Suz.06}.
The cross sections for
the reaction $^{56}$Fe($\nu_e,e^-$)$^{56}$Co were recently analyzed 
to estimate the systematic uncertainty in modeling 
neutrino-nucleus cross sections for medium mass nuclei~\cite{Paa.11}. 
The results were compared with the experimental uncertainty of the 
the KARMEN data. 
The calculated $\nu_e$-$^{56}$Fe cross sections, averaged 
over the Michel spectrum, are listed in Table~1. In addition to the RHB+RQRPA
model calculation based on the density functional DD-ME2, results are shown for
the shell model (GXPF1J) (for $1^+$ transitions) plus the RPA (SGII)
for higher multipoles~\cite{TS.09,Paa.11}, RPA based on the Landau-Migdal
force~\cite{Kol.01},  the QRPA(SIII)\cite{Laz.07} and QRPA based on the G-matrix 
formalism~\cite{Che.10}. By using these methods, the cross section mean 
value and uncertainty are obtained: $<\sigma>_{th}$=(258$\pm$57) $\times$10$^{-42}$cm$^2$,
in excellent agreement with the data from the KARMEN collaboration: 
$<\sigma>_{exp}$=(256$\pm$108$\pm$43) $\times$ 10$^{-42}$cm$^2$. 
At present, the systematic theoretical uncertainty of
the calculated cross section appears to be smaller than the corresponding experimental value.
\begin{table}[pt]
\tbl{Comparison of the inclusive $\nu_e$-$^{56}$Fe cross sections averaged with the Michel flux.}
{\begin{tabular}{@{}cc@{}} \toprule
 & $<\sigma>$ [10$^{-42}$cm$^2$] \\ \colrule
RNEDF (DD-ME2) & 263 \\
SM(GXPF1J) + RPA(SGII)~\cite{TS.09,Paa.11} & 259 \\
RPA (Landau-Migdal)~\cite{Kol.01} & 240 \\
QRPA (SIII)\cite{Laz.07} & 352 \\
QRPA(G-matrix)~\cite{Che.10} & \hphantom{0} 173.5 \\ \colrule
Theoretical average & \hphantom{000}258$\pm$57 \\
Exp. (KARMEN)~\cite{Bod.94} & \hphantom{00000000}256$\pm$108$\pm$43 \\
\botrule
\end{tabular}}
\end{table}

In the environment of high neutrino fluxes that occur in core-collapse supernovae or neutron star mergers, 
neutrino-nucleus reactions are particularly important for the process of nucleosynthesis. 
In addition, neutrino detectors are based on 
inelastic neutrino-nucleus scattering and, to provide reliable
predictions of supernova neutrino induced events, it is essential to 
be able to compute neutrino-nucleus cross sections on a quantitative level.
The supernova neutrino fluxes can be obtained from core-collapse supernova 
simulations~\cite{Fis.12}. However, for the purpose of testing the sensitivity of 
the underlying models of nuclear structure and neutrino-nucleus interactions,
a simplified neutrino flux is employed, described by the Fermi-Dirac 
spectrum,
\begin{equation}
f(E_{\nu}) =  \frac{1}{T^3}  \frac{E_{\nu}^2}
{exp \left[ (E_{\nu}/T)-\alpha  \right ] + 1  } \; ,
\label{fermidirac}
\end{equation}
where T denotes the temperature, and $\alpha$ is the chemical potential. 
As shown in Ref.~\refcite{Paa.13} for the set of inclusive neutrino-nucleus
cross sections calculated using the RNEDF framework, 
the implementation of more realistic fluxes from core-collapse 
supernova simulations is straightforward.

\begin{figure}[th]
\vspace{0.7cm}
\centerline{\psfig{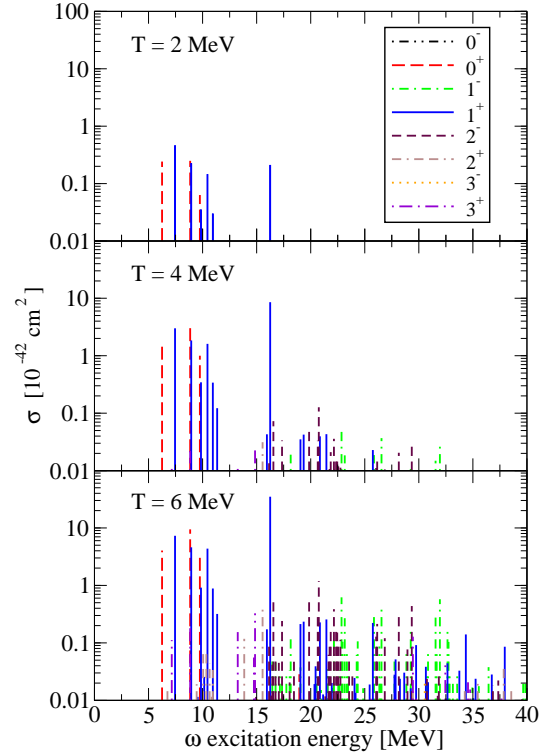}}
\caption{Neutrino-induced excitation spectrum for the reaction $^{56}$Fe($\nu_e,e^-$)$^{56}$Co, 
averaged over the Fermi-Dirac 
distribution Eq.~(\ref{fermidirac}) with $T=2,4$, and 6 MeV, $\alpha =0$. The RHB+RQRPA 
calculations are based on the EDF DD-ME2. The excitation energies are 
given with respect to the ground state of the target nucleus.}
\label{Fe56_neutrino_spectrum}
\end{figure}

\begin{figure}[th]
\vspace{0.7cm}
\centerline{\psfig{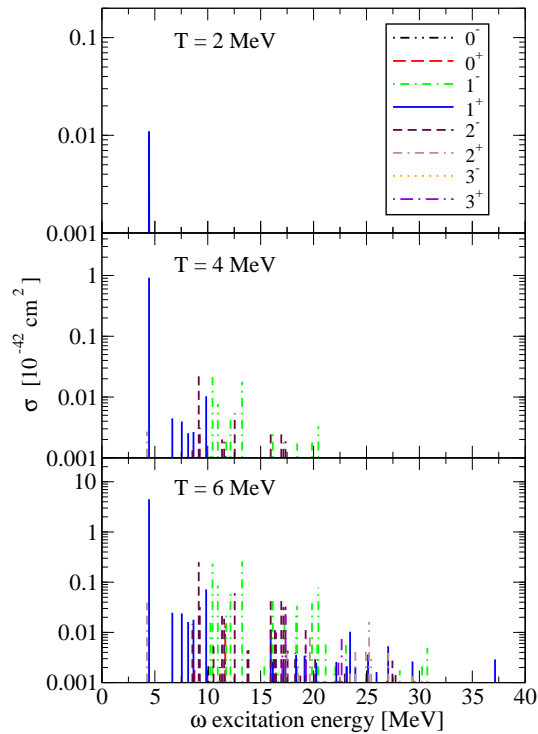}}
\caption{Same as the caption to Fig.~\ref{Fe56_neutrino_spectrum}, but for the antineutrino induced
reactions on $^{56}$Fe.}
\label{Fe56_aneutrino_spectrum}
\end{figure}

For a quantitative analysis of total cross sections, and 
also partial cross sections for neutrino-induced particle knockout, 
in a first step the neutrino-induced excitation spectrum in the daughter 
nucleus has to be computed. In Ref.~\refcite{Kol.01} it has been shown that this spectrum
provides a basis for the implementation of statistical model codes (e.g. SMOKER),
to determine for each final state the branching ratios into various decay channels.
In Fig.~\ref{Fe56_neutrino_spectrum} we plot 
the dominant multipole contributions to the RHB+RQRPA (DD-ME2) 
excitation spectrum induced in the reaction $^{56}$Fe($\nu_e,e^-$)$^{56}$Co. 
The cross sections that correspond to the transition from the ground state of the target nucleus to
each excited state in the daughter nucleus are averaged over the supernova
neutrino flux Eq.~(\ref{fermidirac}), for $T=2,4$, and 6 MeV, and $\alpha =0$. 
Fig.~\ref{Fe56_neutrino_spectrum} shows that, by increasing the temperature
that determines the neutrino flux, the distribution of the neutrino-induced excitation
spectrum is shifted toward higher energies. In addition, at higher temperatures 
the structure of the excitation spectra becomes rather complex due to neutrinos
in the extended tail of the Fermi-Dirac distribution. While for $T=2$ MeV only a few 1$^+$ and 
0$^+$ states contribute to the reaction cross section, at $T=6$ MeV a large
number of states of various multipoles yield non-negligible contributions.
Figure~\ref{Fe56_aneutrino_spectrum} displays the same analysis, but for the
case of inelastic antineutrino -- $^{56}$Fe scattering. For temperatures in the range $T=2-4$ MeV
a single low-lying 1$^+$ state gives the main contribution to the
cross section, whereas at higher temperatures significant contributions arise from 
excited states at higher energies.

\begin{figure}[th]
\centerline{\psfig{file=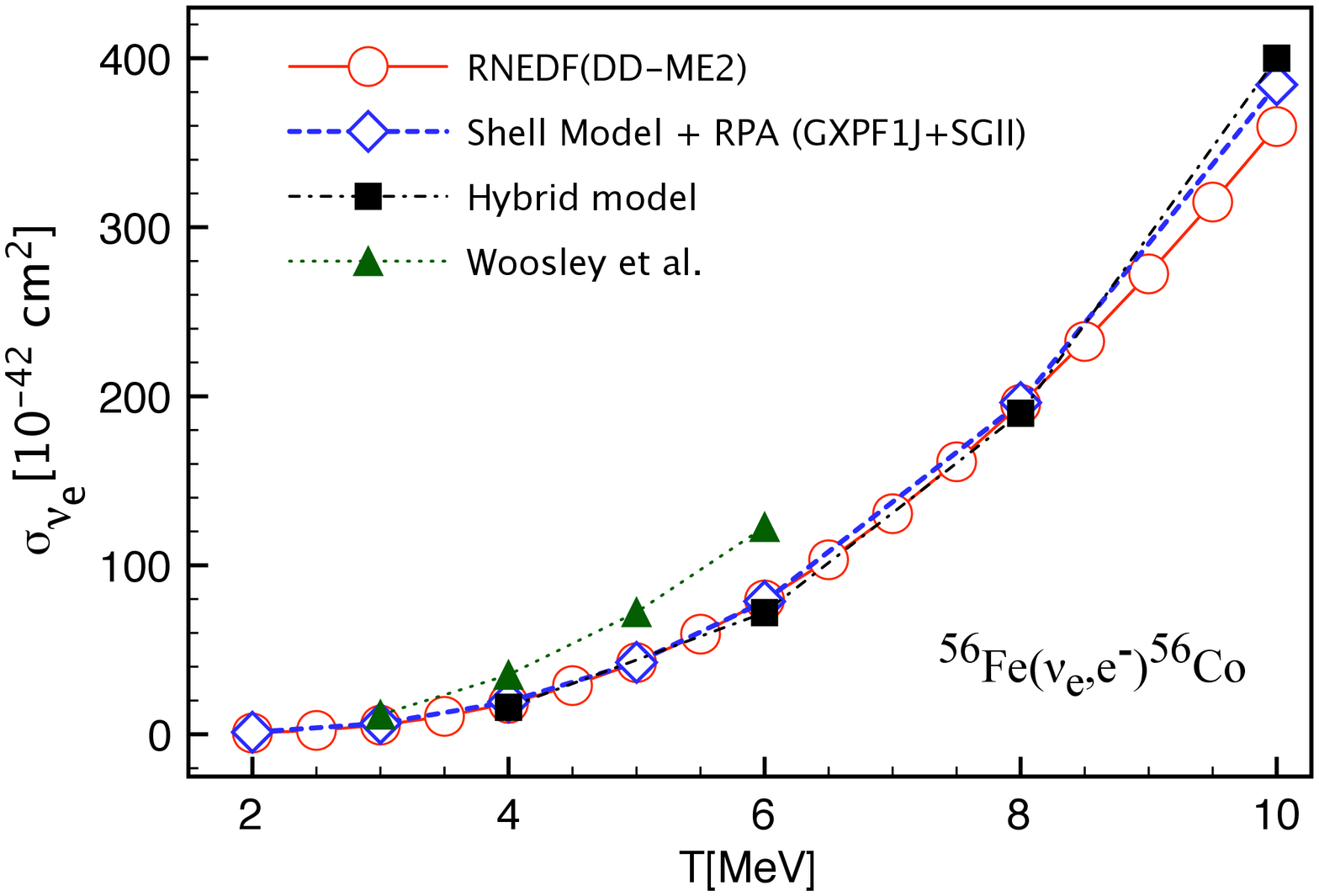,width=9cm,angle=-90}}
\vspace{-1.5cm}
\caption{Inclusive cross sections for the reaction $^{56}$Fe($\nu_e,e^-$)$^{56}$Co,
averaged over the Fermi-Dirac distribution, shown as a function of temperature
in the interval $T=2-10$ MeV. The RHB+RQRPA (DD-ME2) results are compared with
those obtained using the shell model + RPA \cite{TS.09}, and the hybrid model of Ref.~\cite{Kol.01},
as well as those of Ref.~\cite{Woo.90}.}
\label{Fe56_cs_T_Eph200_Jmax5_new}
\end{figure}

The flux-averaged inclusive cross sections for the reaction $^{56}$Fe($\nu_e,e^-$)$^{56}$Co, 
evaluated at different temperatures in the interval $T= 2 - 10$ MeV, $\alpha=0$, 
are displayed in Fig.~\ref{Fe56_cs_T_Eph200_Jmax5_new}.
The results of RHB+RQRPA (DD-ME2)
calculations are compared to those obtained with the shell 
model + RPA (GXPF1J + SGII)~\cite{TS.09}, the hybrid model of Ref.~\cite{Kol.01}, and 
estimates from Ref.~\refcite{Woo.90} (Woosley et al.). As shown in the figure, an excellent
agreement is obtained for the RHB+RQRPA, shell model and hybrid model,
with small deviations noticeable only at the high-end temperature $T=10$ MeV. At 
temperatures above $T=4$ MeV the results of Ref.~\refcite{Woo.90} display deviations 
with respect to more recent microscopic calculations.

\section{\label{sec6}Large-scale calculations of neutrino-nucleus reactions}

\begin{figure}[th]
\centerline{\psfig{file=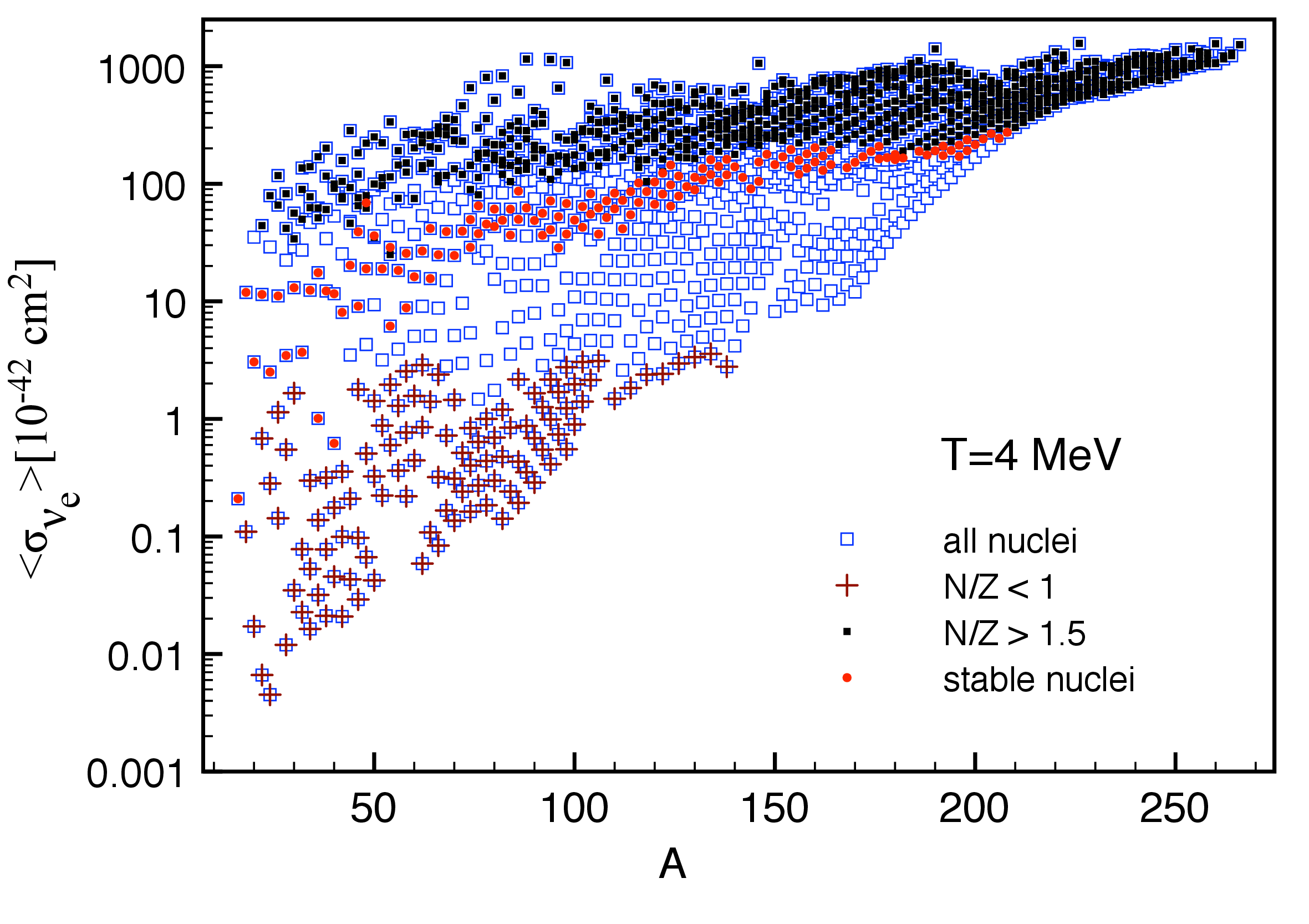,width=11cm}}
\caption{RHB+RQRPA (DD-ME2) inclusive neutrino-nucleus cross sections
averaged over the Fermi-Dirac distribution for $T=4$ MeV, $\alpha =0$, 
as a function of the mass number of target nuclei. The cross sections for 
particular groups of target nuclei are further marked with filled (red) 
circles for stable nuclei, crosses for nuclei with $N/Z < 1$, and 
filled (blue) squares for nuclei with $N/Z > 1.5$.}
\label{all_isotopes_T4}
\end{figure}
Inelastic neutrino-nucleus scattering plays an important role in supernova evolution and nucleosynthesis~\cite{Jan.07}. During the star collapse 
phase, charged-current $(\nu_e,e^-)$ reactions on nuclei are suppressed because of Pauli blocking of the final 
electron phase space~\cite{Bru.91}.
However, in the environment of exploding massive stars or neutron star mergers, considered as 
possible sites for the r-process nucleosynthesis, charged-current neutrino-nucleus reactions 
could contribute to the production of heavier isotopes. Since a complete modeling of neutrino-induced reactions requires the inclusion of not only Gamow-Teller type of transitions but also contributions from higher multipoles, including a large pool of target nuclei 
can be computationally very demanding. Even though 
the shell model provides an accurate description of ground state correlations and low-lying excited states, 
the calculation of high-lying excitations and/or higher multipole can be a formidable task, making 
this approach applicable essentially only to allowed transitions in light and medium-mass nuclei. 
For systematic studies of neutrino-nucleus cross sections over the entire nuclide chart, 
microscopic calculations must therefore be implemented using a framework based on universal energy
density functionals.

A consistent set of models based on the RNEDF framework was
recently employed in large-scale calculations of charged-current neutrino-nucleus cross sections, for a set of even-even target nuclei from oxygen to lead ($Z=8-82$), and with the neutron number spanning the interval $N=8-182$~\cite{Paa.13}. These extensive calculations include allowed and forbidden transitions of multipolarity $J^{\pi}=0^{\pm} - 5^{\pm}$. The inclusive cross sections and a full set of flux-averaged cross sections for the $Z=8-82$ pool of nuclei is available for the range of temperatures $T=0-10$ MeV~\cite{Paa.13}.
As an illustrative example, the $(\nu_e,e^-)$ cross sections, averaged over the supernova neutrino flux (\ref{fermidirac}) for  $T=4$ MeV, 
$\alpha=0$ are shown in Fig.~\ref{all_isotopes_T4}. 
The cross sections are additionally marked for different groups of target nuclei: 
(i) stable nuclei, (ii) nuclei with proton excess $N/Z < 1$, and (iii) neutron-rich
nuclei with with $N/Z > 1.5$. In the case of neutron-rich nuclei the
cross sections are enhanced (blue squares) in comparison to stable nuclei (up to an order of magnitude), 
because of the larger number of neutrons that can participate in charge-exchange neutrino-induced reactions. 
On the other side of the valley of stability, for neutron-deficient and proton-rich nuclei, 
the $(\nu_e,e^-)$ cross sections are considerably reduced (crosses)
due to the Pauli blocking of proton orbitals available for neutrino-induced transitions.

\begin{figure}[th]
\centerline{\psfig{file=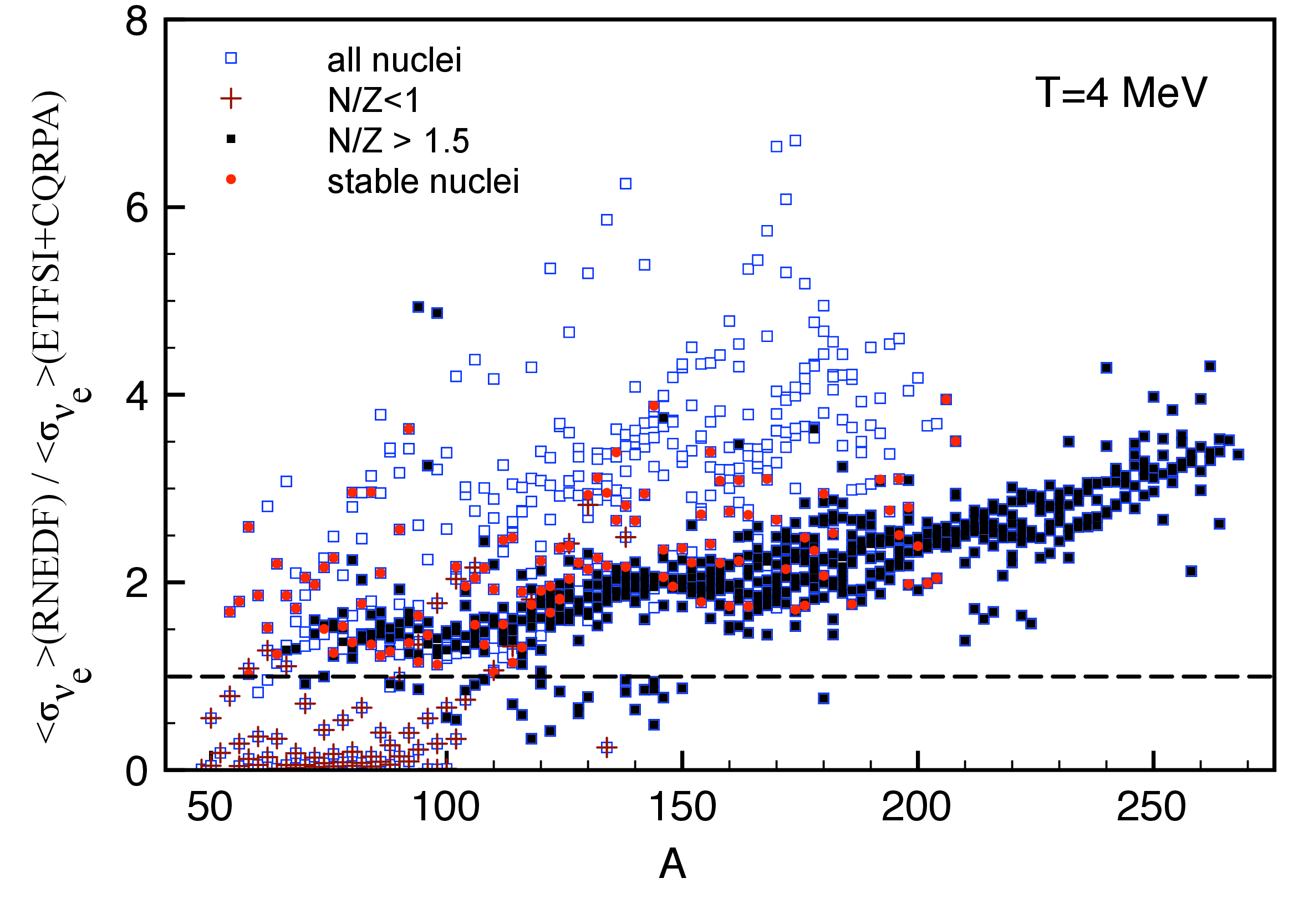,width=11cm}}
\caption{The ratio of the RHB+RQRPA cross sections shown in Fig.~\ref{all_isotopes_T4} and the cross sections
calculated using the ETFSI+CQRPA~\protect\cite{Bor.00} framework.}
\label{all_isotopes_T4_RQRPA_Goriely.eps}
\end{figure}
The calculated neutrino-nucleus cross sections can be used for modeling the detector response to neutrinos
involving different target nuclei, as well as in r-process calculations involving various astrophysical
scenarios. It is, therefore, important to determine systematic uncertainties in the 
calculation of neutrino-nucleus cross sections when various theoretical frameworks and effective interactions 
are employed. For the example of large-scale modeling of cross sections averaged over the supernova neutrino 
flux for T=4, $\alpha=0$, the results of self-consistent calculations based on the RNEDF 
framework have been compared to those obtained using the RPA (WS+LM)~\protect\cite{Lan.01} 
and ETFSI+CQRPA~\protect\cite{Bor.00}. Figure~\ref{all_isotopes_T4_RQRPA_Goriely.eps} displays the ratio of supernova neutrino flux-averaged cross sections calculated with the RHB+RQRPA (DD-ME2) and ETFSI+CQRPA~\cite{Bor.00}, as a function of the mass number of target nuclei. 
To identify differences specific to particular groups of nuclei, the results are additionally labeled   
for stable nuclei, neutron-deficient nuclei with $N/Z<1$, and neutron-rich nuclei with $N/Z > 1.5$.
The RHB+RQRPA cross sections include higher multipoles up to $J=5$, whereas the ETFSI+CQRPA results contain 
only IAS and GT transitions. For most nuclei the RHB+RQRPA approach based on the RNEDF framework 
predicts systematically larger cross sections, up to a factor of $\approx 4-6$. 
In the case of neutron rich-nuclei ($N/Z > 1.5$), for which the
absolute values of the cross sections are significant, the differences between the two models behave in 
rather systematic way and the ratio of the cross sections increases with an 
almost linear mass dependence. In the mass region $A= 50-100$, for the set of $N/Z<1$ nuclei
the RHB+RQRPA cross sections are smaller than the corresponding ETFSI+CQRPA ones.
The reason is to a large extent the anomalous enhancement of the cross 
sections for neutron-deficient nuclei predicted by the ETFSI+CQRPA model (cf. Ref.~\refcite{Paa.13} for more details).

\section{\label{sec7}Beta-decay half-lives}
An important application of elementary charge-exchange transition studies is the evaluation of $\beta$-decay rates 
of neutron-rich nuclei. In the $\beta^{-}$ decay process  
\begin{equation}
^{A}_{Z}X_{N} \to \, ^{A}_{Z+1}Y_{N-1} + e^{-} + \bar{\nu}
\end{equation}
a neutron is transformed into a proton, and this additional proton enhances the Coulomb repulsion pushing the isobaric analogue state above the $Q_{\beta}$ energy window. Fermi transitions, therefore, do not contribute to the decay of neutron-rich nuclei, and it is the low-energy part of the 
Gamow-Teller spectrum that determines the decay rate~\cite{Behrens1969}
\begin{equation}
\lambda_{i} = D \int_{1}^{W_{0,i}} W \sqrt{W^{2} - 1} \left(W_{0,i} - W\right)^{2} F(Z,W) C(W) dW.
\end{equation}
Here, $W$ is the emitted electron energy in units of the electron mass, $W_{0,i}$ is the maximum electron energy, $F(Z,W)$ is the Fermi function that corrects for the Coulomb field of the nucleus. $C(W)$ is the \emph{shape factor} which, when only allowed transitions are taken into account, 
depends only on the Gamow-Teller transition strength. In neutron-rich nuclei, however, high-lying neutron states are partially occupied enabling first-forbidden, parity-changing transitions, such as the spin-dipole, to provide a significant contribution to the total decay rate. In that case the shape factor becomes a complicated combination of transition matrix elements and terms that include the energy dependence~\cite{Behrens1971,Behrens1982,Zhi2013}. 

As the contribution of the phase space available to leptons depends on the energy of the transition, for the evaluation of decay rates it is important to achieve a good description of the low-energy part of the spectrum. Thus, the $T = 0$ pairing interaction Eq. (\ref{pn-pair}) is an essential ingredient for  any self-consistent mean-field model that describes the decay properties of neutron-rich nuclei~\cite{Eng.99}. In several previous studies it was found that a single value of the pairing strength cannot be used for all isotopic chains or in different mass regions~\cite{Eng.99,Nik.05,Mar.07}. However, with the assumption that the strength of the $T = 0$ pairing depends on the difference between the number of neutrons and protons~\cite{Niu2013} 
\begin{equation}
V_{0} = V_{L} + \frac{V_{D}}{1 + e^{a + b(N-Z)}},
\end{equation}
it is possible to obtain a very good agreement with data over the entire nuclear chart. The results presented in this 
section have been obtained using the values: $V_{L} = 160.0$~MeV, $V_{D} = 15.0$ MeV, $a = 7.2$ and $b= -0.3$, adjusted to obtain the best possible description of available data on $\beta$-decay
half-lifes ~\cite{Audi2012}.

\begin{figure}[htb]
\centerline{\psfig{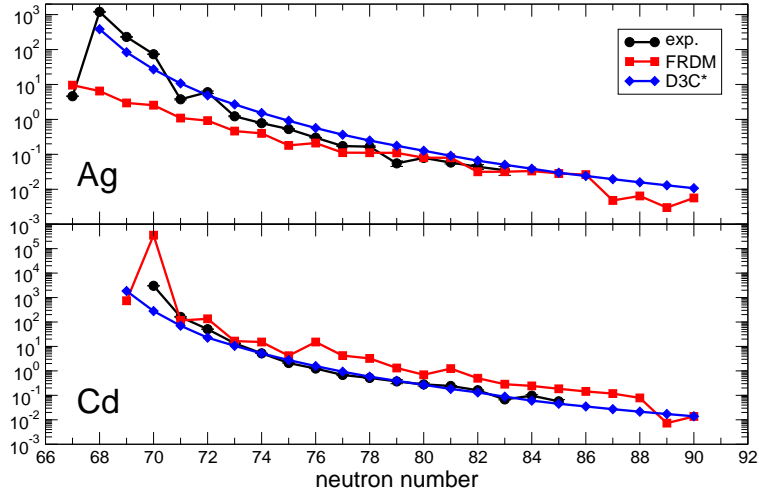}}
\caption{\label{fig:thalf} $\beta$-decay half-lives of neutron-rich isotopes of silver (top) and cadmium (bottom), obtained with two theoretical approaches and compared to experimental values. The FRDM+QRPA values are denoted by (red) squares, RHB+RQRPA (D3C*) results by (blue) diamonds. 
For most nuclei the error bars on data points are too small to be discernible.}
\end{figure}

In Fig.~\ref{fig:thalf} we display $\beta$-decay half-lives calculated with the finite-range droplet model (FRDM) + QRPA~\cite{Moeller2003} (red squares) and the microscopic RHB+RQRPA model (effective interaction D3C*)~\cite{Mar.07} (blue diamonds), 
in comparison to available data~\cite{Audi2012}. In the top panel results for the isotopes of silver ($Z = 47$) are shown for $N = 67-90$.  A prominent feature is the relatively large discrepancy between model predictions for nuclei close to stability. The RHB+RQRPA (D3C*) results reproduce the empirical half-lives and follow closely the decreasing trend of the data, except in the case of $^{114}$Ag. The half-life of this nucleus is shorter by two orders of magnitude compared to neighboring isotopes, and this is very difficult to reproduce in model calculations. The FRDM reproduces the decay rate of $^{114}$Ag but systematically underestimates the decay half-lives of longer-lived isotopes. At the very neutron-rich side both models predict results in good agreement with data. We notice, however, that the FRDM results display a sudden deviation from the general trend (between $N = 86$ and $N = 90$) by as much as an order of magnitude, while the D3C*-based RHB+RQRPA predicts a very smooth mass dependence of the decay half-lives.

In the case of the cadmium isotopic chain ($Z = 48$), the predictions of the two models are in better agreement and, generally, both models 
reproduce the data. Systematic deviations from the experimental values and RHB+RQRPA results are apparent in the FRDM predictions for the isotopes $N \ge 76$. One may notice that for both silver and cadmium isotopes the FRDM results display an odd-even staggering, not observed in the data. This staggering and the systematic deviation from the data may have important consequences in r-process simulations and the resulting elemental abundances. 

In both isotopic chains the main contribution to the decay rate originates from the very strong Gamow-Teller transition $\nu g_{7/2} \to \pi g_{9/2}$, especially for nuclei closer to stability. With the addition of more neutrons, in particular above the $N = 82$ closed neutron shell, additional configurations become available and first-forbidden transitions produce a significant contribution to the total decay rate. 
\begin{figure}[htb]
\centerline{\psfig{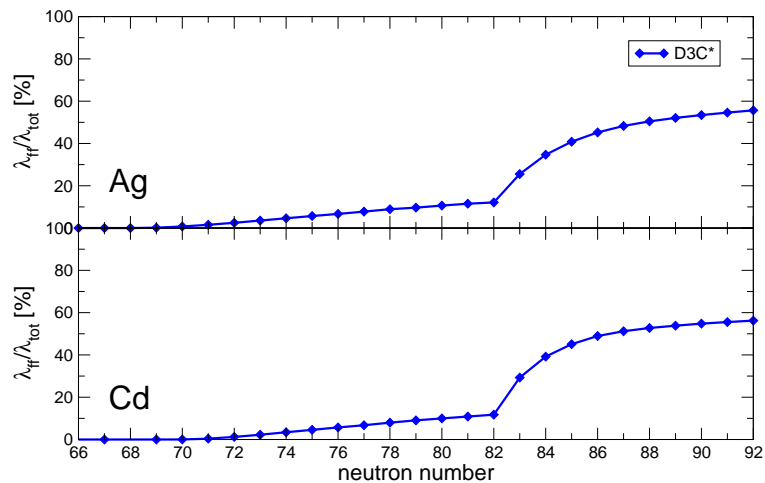}}
\caption{\label{fig:ff} Contribution (in percentage) of first-forbidden transitions to the total $\beta$-decay rates of silver and cadmium isotopes.}
\end{figure}
If Fig~\ref{fig:ff} we display the contribution of first-forbidden transitions to the total decay rate of silver and cadmium isotopes, from $N = 66$ to $N = 92$. Below the $N = 82$ closed neutron shell the contribution of first-forbidden transitions is rather small (up to 10\%), and limited to the $J^{\pi} = 1^{-}$ $\, \nu h_{11/2} \to \pi g_{9/2}$ configuration. Above the closed shell, however, the neutron orbital $f_{7/2}$ and, to a lesser extent $h_{9/2}$, become partially occupied and contribute to both the $0^{-}$ and the $1^{-}$ transitions. The occupation of opposite parity states in the next neutron shell 
leads to the rapid increase of the contributions shown in Fig.~\ref{fig:ff} above the neutron shell closure. 

An important property of $\beta$-unstable nuclei is the number of neutrons emitted after the decay, that is, the $\beta$-delayed neutron emission probability. The difference between the time scales of the $\beta$-decay and the subsequent particle emission process justifies the assumption of their statistical independence. Thus, the $\beta$-delayed neutron emission is considered as a multi-step process that proceeds starting with the $\beta$-decay of the parent nucleus $(A,Z)$, followed by either the $\gamma$ de-excitation of the daughter nucleus $(A,Z+1)$, or by the emission of a neutron and the formation of the final nucleus $(A-1,Z+1)$. 
The probability of emission of a single or multiple neutrons can be
expressed ~\cite{Pappas1972}:
\begin{equation} \label{eq:sep}
P_{xn} = \frac{\sum_{S_{xn} \le E_{i} \le S_{(x+1)n}}
\lambda_{i}}{\lambda_{tot}}\;,
\end{equation}
with the assumption that the daughter nucleus will emit as many
neutrons as energetically allowed. Here, $\lambda_{i}$ is the partial
decay rate, $\lambda_{tot}$ the total decay rate, and the sum takes into
account all transitions with energies between $x$ and $(x+1)$ neutron
separation energies.
\begin{figure}[htb]
\centerline{\psfig{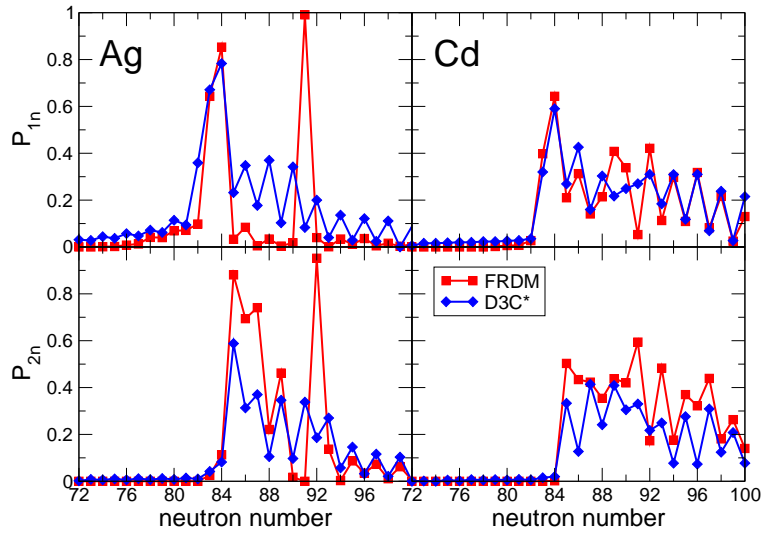}}
\caption{\label{fig:emission} 
Comparison of beta-delayed neutron emission probabilities calculated in 
the FRDM + QRPA (red squares) and RHB+RQRPA (D3C*) (blue
diamonds) models. The top panels display the one-neutron emission probabilities
$P_{1n}$ for the silver (left) and cadmium (right) isotopic chains, while
in the bottom panels the two-neutron emission probabilities $P_{2n}$ are shown.
}
\end{figure}

In Fig.~\ref{fig:emission} we compare the one ($P_{1n}$) and two ($P_{2n}$) neutron emission probabilities for silver and cadmium isotopes. The odd-even staggering of the probabilities is a direct consequence of the staggering of the neutron separation energies in the daughter nuclei. A nucleus with an even number of neutrons decays into a nucleus with an odd number of neutrons, i.e. the one-neutron separation energy $S_{n}$ is smaller in the daughter nucleus. In the case of a parent with an odd neutron number, the decay is into a daughter nucleus with an even number of neutrons and a correspondingly larger $S_{n}$. For a parent with an even number of neutrons the sum in Eq. (\ref{eq:sep}) takes into account a wider energy interval and this results in a higher neutron emission probability. 

For the lighter isotopes of both chains, the two models (FRDM + QRPA (red squares) and RHB+RQRPA (D3C*) (blue
diamonds)) predict a similar dominance of the one-neutron emission up to $N = 84$. Significant differences are found above
$N = 84$ in silver isotopes. 
The FRDM predicts very small values of $P_{1n}$, and comparatively much larger values of $P_{2n}$ for nuclei up to $N = 92$, whereas 
the RHB+RQRPA (D3C*) yields a more gradual decrease of both $P_{1n}$ and $P_{2n}$. Above $N = 92$ both models give small 
values of $P_{1n}$ and $P_{2n}$, because for very neutron-rich nuclei the emission of 3 or more neutrons dominates. The differences between 
model predictions are much less pronounced in the case of Cd isotopes, especially for the two-neutron emission probability.

\section{\label{sec8}Conclusions}

Weak-interaction processes provide valuable information on ground-state properties and the structure of excited states of a large number 
of nuclei that lie on both sides of the valley of $\beta$-stability. Calculations of stellar nucleosynthesis, nuclear aspects of supernova collapse 
and explosion, and neutrino-induced reactions, necessitate as input properties of thousands of nuclei far from stability, including characteristics  
related to weak-interaction processes. Many of these nuclei, especially on the neutron-rich side, are not accessible in experiments and their 
properties, therefore, must be determined using different theoretical approaches. One of these is the self-consistent mean-field method based 
on universal nuclear energy density functionals or global effective interactions, which can be applied in the description of arbitrarily heavy nuclei, 
exotic nuclei far from stability, and even nuclei at the nucleon drip-lines. 

We have reviewed some recent applications of a theoretical framework based on relativistic energy density functionals in modeling neutrino-nucleus 
reactions and calculation of $\beta$-decay rates. Starting from a self-consistent mean-field solution for the nuclear ground state, a consistent 
proton-neutron relativistic quasiparticle random-phase approximation is used to calculate collective excited states in daughter nuclei and compute 
cross sections and transition rates. The model has very successfully been applied not only to the weak-interaction processes described in the present 
article, but also to muon capture and stellar electron capture~\cite{Mar.09,Niu.11}. The principal advantage of using this approach is the ability to perform systematic 
calculations of weak-interaction rates in nuclei of arbitrary mass 
along the $r$-process path, based on microscopic self-consistent mean-field potentials rather than 
empirical potentials, and the inclusion of transitions of arbitrary multipolarity. Future developments will include the exploration of more complex 
configurations, e.g. particle-vibration coupling effects, a systematic treatment of finite-temperature effects, and the extension of the 
RHB+RQRPA model to explicitly take into account quadrupole deformations.

\section*{Acknowledgements}

This work has been supported in part by Deutscher Akademischer Austausch Dienst (DAAD), the Marie Curie FP7-PEOPLE-2011-COFUND program NEWFELPRO, the Swiss National Science Foundation and by the IAEA Research Contract No. 18094/R0.

\end{document}